\input{aipcheck}

\documentclass[
    ,final            
  ]
  {aipproc}

\layoutstyle{6x9}

\def\beq{\begin{equation}}
\def\ee{\end{equation}}
\def\eeq{\end{equation}}

\def\bfig{\begin{figure}}
\def\efig{\end{figure}}
\def\bea{\begin{eqnarray}}
\def\beann{\begin{eqnarray*}}
\def\eea{\end{eqnarray}}
\def\eeann{\end{eqnarray*}}

\def\raw{\rightarrow}

\usepackage{graphicx}

\begin{document}

\title{Meson decay in the Fock-Tani Formalism}



\author{D.T. da Silva}{
  address={Instituto de F\'{\i}sica, Universidade Federal do Rio Grande do Sul,
 Caixa Postal 15051, CEP 91501-970, Porto Alegre, RS, Brazil.}
}

\author{M. L. L. da Silva}{
  address={Instituto de F\'{\i}sica, Universidade Federal do Rio Grande do Sul, Caixa Postal 15051, CEP 91501-970, Porto Alegre, RS, Brazil.}
}

\author{D. Hadjimichef}{
  address={Departamento de F\'{\i}sica, Instituto de F\'{\i}sica e Matem\'atica, 
Universidade Federal de Pelotas, 
Campus Universit\'ario, CEP 96010-900, Pelotas, RS, Brazil.},
altaddress={Instituto de F\'{\i}sica, Universidade Federal do Rio Grande do Sul, 
Caixa Postal 15051, CEP 91501-970, Porto Alegre, RS, Brazil.}
}

\author{C. A. Z. Vasconcellos}
{
address={Instituto de F\'{\i}sica, Universidade Federal do Rio Grande do Sul, 
Caixa Postal 15051, CEP 91501-970, Porto Alegre, RS, Brazil.}
}

\begin{abstract}

    The Fock-Tani formalism is a first principle method to obtain effective interactions
from microscopic Hamiltonians. Usually this formalism was applied to scattering, here
we introduced it to calculate  partial decay widths for  mesons. 

\end{abstract}

\maketitle


\section{Introduction}

  In the Fock-Tani representation \cite{annals} one starts with the Fock representation of the
  system using field operators of elementary constituents which satisfy canonical
  (anti) commu\-ta\-tion relations. Composite-particle field operators  are 
  linear combinations of the elementary-particle operators and do not generally
  satisfy canonical (anti) commutation relations. ``Ideal" field 
  operators acting on an enlarged Fock space are then introduced in close 
  correspondence with the composite ones. The enlarged Fock space is a graded 
  direct product of the original Fock space and an ``ideal state space". The 
  ideal operators correspond to particles with the same quantum numbers of the 
  composites; however, they satisfy by definition canonical (anti)commutation 
  relations. Next, a given unitary transformation, which transforms the 
  single composite states into single ideal states, is introduced. When the 
  transformation acts on operators in the subspace of the enlarged Fock space 
  which contains no ideal particles, the transformed operators explicitly 
  express the interactions of composites and constituents. Application of the 
  unitary operator on the microscopic Hamiltonian, or on 
  other hermitian operators expressed in terms of the elementary constituent 
  field operators, gives equivalent operators 
  which contain the ideal field operators.
  In the present we apply the formalism to meson decay into two-mesons. In particular
  the example system is $\rho^{+}\rightarrow \pi^{+}\pi^{0}$.

\section{The Fock-Tani formalism}

Now let us to apply the Fock-Tani formalism in the microscopic Hamiltonian
to obtain an effective Hamiltonian. In the Fock-Tani formalism we can write
the meson creation operators in the following form 
\begin{eqnarray*}
\,\,\,\,M_\alpha ^{\dagger }=\Phi _\alpha ^{\mu \nu }q_\mu ^{\dagger }\bar{q}%
_\nu ^{\dagger }.  \label{G}
\end{eqnarray*}
where $\,$ $\Phi _\alpha ^{\mu \nu }$ is the bound-state wave-functions for
two-quarks respectively.The operators of the quark and antiquark obey the
following anticommutation relations 
\begin{eqnarray*}
&&\{q_\mu ,q_\nu \}=\{\bar{q}_\mu ,\bar{q}_\nu \}=\{q_\mu ,\bar{q}_\nu
\}=\{q_\mu ,\bar{q}_\nu ^{\dag }\}=0  \nonumber \\
&&\{q_\mu ,q_\nu ^{\dagger }\}=\{\bar{q}_\mu ,\bar{q}_\nu ^{\dagger
}\}=\delta _{\mu \nu }.
\end{eqnarray*}
The composite meson operators satisfy non-canonical commutation relations 
\begin{eqnarray*}
&& \\
\,\,\,\,\,[M_\alpha ,M_\beta ^{}] &=&0\,;\,\,\,\,[M_\alpha ,M_\beta
^{\dagger }]=\delta _{\alpha \beta }-\Delta _{\alpha \beta }
\end{eqnarray*}
where 
\begin{eqnarray*}
&&\,\,\,\Delta _{\alpha \beta }=\Phi _\alpha ^{\star \mu \gamma }\Phi _\beta
^{\gamma \rho }q_\rho ^{\dagger }q_\mu +\Phi _\alpha ^{\star \mu \gamma
}\Phi _\beta ^{\gamma \rho }\bar{q}_\rho ^{\dagger }\bar{q}_\mu .
\end{eqnarray*}
The idea of the Fock-Tani formalism is to make a representation change, of
form that the composite particles operators are described by operators that
satisfy canonical commutation relations. i.e., which obey canonical
relations 
\begin{eqnarray*}
&&[m_\alpha ,m_\beta ]=0\,\,\,\,\,;\,\,\,\,[m_\alpha ,m_\beta ^{\dagger
}]=\delta _{\alpha \beta }.
\end{eqnarray*}
where $m_\alpha ^{\dagger }$ is the operator of the ``ideal particle''
creation. This way one can transform the composite state $|\alpha \rangle $
into an ideal state $|\alpha \,)$, in the meson case for example we have 
\begin{eqnarray*}
U^{-1}\,M_\alpha ^{\dagger }\,\left| 0\right) =m_\alpha ^{\dagger }\left|
0\right) =|\alpha \,)
\end{eqnarray*}
where $U=\exp ({tF})$ and $F$ is the generator of the meson transformation
given by 
\begin{eqnarray*}
F=\sum_\alpha m_\alpha ^{\dag }\tilde{M}_\alpha -\tilde{M}_\alpha ^{\dag
}m_\alpha  \label{F}
\end{eqnarray*}
with 
\begin{eqnarray*}
\tilde{M}_\alpha =M_\alpha +\frac 12\Delta _{\alpha \beta }M_\beta +\frac
12M_\beta ^{\dagger }[\Delta _{\beta \gamma },M_\alpha ]M_\gamma .
\end{eqnarray*}
A set of Heisenberg-like equations for the basic operators $m,M,q$ 
\begin{eqnarray*}
\frac{dm_\alpha (t)}{dt}=[m_\alpha (t),F]=\tilde{M}_\alpha
(t)\,\,\,\,\,;\,\,\,\,\,\frac{d\tilde{M}_\alpha (t)}{dt}=[\tilde{M}_\alpha
(t),F]=-m_\alpha (t)\,.
\end{eqnarray*}
The simplicity of these equations are not present in the equations for $q$
\[
\frac{dq_\mu (t)}{dt}=[q_\mu (t),F]=-\Phi _\alpha ^{\mu \nu }\bar{q}%
_\nu ^{\dagger }(t)m_\alpha (t) 
\]
and
\[
\frac{d\bar{q}_\nu ^{}(t)}{dt}=[\bar{q}_\nu ^{}(t),F]=\Phi _\alpha
^{\mu \nu }q_\mu ^{\dagger }(t)m_\alpha (t). 
\]
The solution for these equations can be found order by order in the
wavefunctions. For zero order one has $q_\mu ^{(0)}(t)=q_\mu $, $\bar{q}_\nu
^{(0)}(t)=\bar{q}_\nu $, $m_\alpha ^{(0)}(t) =M_\alpha \sin {t}+m_\alpha \cos {t} $
$M_\beta ^{(0)}(t) =M_\beta \cos {t}-m_\beta \sin {t}$.
 In the first order $m_\alpha ^{(1)}=0,\,\,\,M_\beta ^{(1)}=0$ and 
$q_\mu ^{(1)}(t)=-\Phi _\alpha ^{\mu \nu _1}\bar{q}_{\nu _1}^{\dagger
}(M_\alpha -m_\alpha ) { }$; $\bar{q}_\nu ^{(1)}(t)=\Phi _\alpha
^{\mu _1\nu }q_{\mu _1}^{\dagger }(M_\alpha -m_\alpha )$.
for these calculations, we are using $t=-\frac \pi 2$
Consequently, for second and third order, one obtains 
\beann
q_\mu ^{(2)}(t)&=&\frac 12\Phi _\alpha ^{*\mu _2\nu _1}\Phi _\beta ^{\mu \nu
_1}[-m_\alpha ^{\dagger }m_\beta -M_\alpha ^{\dagger _{}}M_\beta +M_\alpha
^{\dagger }m_\beta ^{}] { }q_{\mu _2}^{} { } 
\\
\bar{q}_\nu ^{(2)}(t)&=&\frac 12\Phi _\alpha ^{*\mu _1\nu _2}\Phi _\beta ^{\mu
_1\nu }[-m_\alpha ^{\dagger }m_\beta -M_\alpha ^{\dagger _{}}M_\beta
+M_\alpha ^{\dagger }m_\beta ^{}] { }\bar{q}_{\nu _2}^{} 
\eeann
and
\begin{eqnarray*}
q_\mu ^{(3)}(t) &=&\frac 12\Phi _\alpha ^{*\rho \sigma }\Phi _\beta ^{\mu
\sigma }\Phi _\gamma ^{\rho \sigma _1}\bar{q}_{\sigma _1}^{\dagger
}[-m_\alpha ^{\dagger }m_\beta m_\gamma -M_\alpha ^{\dagger _{}}m_\beta
M_\gamma +m_\alpha ^{\dagger }m_\beta M_\gamma ^{} \\
&&+M_\alpha ^{\dagger _{}}M_\beta M_\gamma ] { }+\frac 12\Phi _\alpha
^{*\rho \sigma }\Phi _\alpha ^{\mu \nu _1}\Phi _\beta ^{\rho \sigma _1}\bar{q%
}_{\nu _1}^{\dagger }\bar{q}_{\sigma _1}^{\dagger }\bar{q}_\sigma
^{}[M_\beta +m_\beta ^{}] \\
&&+\frac 12\Phi _\alpha ^{*\rho \sigma }\Phi _\alpha ^{\mu \nu _1}\Phi
_\beta ^{\rho _1\sigma _{}}\bar{q}_{\nu _1}^{\dagger }q_{\rho _1}^{\dagger
}q_\rho ^{}[M_\beta +m_\beta ^{}]
\\
\bar{q}_\nu ^{(3)}(t) &=&\frac 12\Phi _\alpha ^{*\rho \sigma }\Phi _\beta
^{\rho \nu }\Phi _\gamma ^{\rho _1\sigma }q_{\rho _1}^{\dagger }[-m_\alpha
^{\dagger }m_\beta m_\gamma -M_\alpha ^{\dagger _{}}m_\beta M_\gamma
+m_\alpha ^{\dagger }m_\beta M_\gamma ^{} \\
&&+M_\alpha ^{\dagger _{}}M_\beta M_\gamma ] { }+\frac 12\Phi _\alpha
^{*\rho \sigma }\Phi _\alpha ^{\mu _1\nu }\Phi _\beta ^{\rho \sigma
_1}q_{\mu _1}^{\dagger }\bar{q}_{\sigma _1}^{\dagger }\bar{q}_\sigma
^{}[M_\beta +m_\beta ^{}] \\
&&+\frac 12\Phi _\alpha ^{*\rho \sigma }\Phi _\alpha ^{\mu _1\nu }\Phi
_\beta ^{\rho _1\sigma _{}}q_{\mu _1}^{\dagger }q_{\rho _1}^{\dagger }q_\rho
^{}[M_\beta +m_\beta ^{}]
\end{eqnarray*}

\section{The Microscopic Model}

In this model we use the following Hamiltonian inspired in the $^{3}P_{0}$ approach
\begin{eqnarray*}
H=g\int d^3x\Psi ^{\dag }(\vec{x})\gamma ^0\Psi (\vec{x}).
\end{eqnarray*}
Where the quark field is respectively 
\begin{eqnarray*}
\Psi (\vec{x})=\sum_s\int \frac{d^3p}{(2\pi )^{3/2}}[u(\stackrel{\rightarrow 
}{p},s)b(\stackrel{\rightarrow }{p},s)+v(-\stackrel{\rightarrow }{p}%
,s)d^{\dag }(-\stackrel{\rightarrow }{p},s)]e^{i\stackrel{\rightarrow }{p}%
\cdot \vec{x}}.
\end{eqnarray*}
and $\gamma =\frac g{2m_q}$
where $m_q$ is the mass of both produced quarks. 
After momentum the expansion of the quark field and retaining only
the $b^{\dagger}d^{\dagger}$ yields
\beann
H_{I}=g\sum_{ss^{\prime}}\int d^{3}p\, d^{3}p^{\prime}\delta
\left(\overrightarrow{p}+\overrightarrow{p}\,\,^{\prime}\right)u_{s^{\prime}}^{\dagger}
\left(\overrightarrow{p}\,\,^{\prime}\right)\gamma^{0}\upsilon_{s}
\left(\overrightarrow{p}\,\,\right)b_{s^{\prime}}^{\dagger}
\left(\overrightarrow{p}\,\,^{\prime}\right)d_{s}^{\dagger}
\left(\overrightarrow{p}{}\right).
\eeann
Introducing the following notation $b\rightarrow q$; $d\rightarrow\bar{q}$;
$\mu=\left(\overrightarrow{p}\,\,^{\prime},\, s^{\prime}\right)$;
$\upsilon=\left(\overrightarrow{p},\, s\right)$ and
\beann
V_{\mu\nu}=g\sum_{ss^{\prime}}\int d^{3}p\, d^{3}p^{\prime}\delta\left
(\overrightarrow{p}+\overrightarrow{p}\,\,^{\prime}\right)u_{s^{\prime}}^{\dagger}
\left(\overrightarrow{p}\,\,^{\prime}\right)\gamma^{0}\upsilon_{s}
\left(\overrightarrow{p}\,\,\right).
\eeann
The Hamiltonian now is reduced to a compact form,
$H_{I}=V_{\mu\nu}\,\,q_{\mu}^{\dagger}\bar{q}_{\nu}^{\dagger}$
where sum (integration) is implied over repeated indexes. Applying
the Fock-Tani transformation to $H_{I}$ one obtains the effective
Hamiltonian $H_{I_{FT}}=U^{-1}H_{I}U$. 
For a $q\bar{q}$ meson $A$ to decay to mesons $B+C$ we must have
$
\left(q\bar{q}\right)_{A}\rightarrow\left(q\bar{q}\right)_{B}+\left(q\bar{q}\right)_{C}
$ .
The transformed operators that shall contribute to $U^{-1}H_{I}U$
are\[
H_{m}=V_{\mu\nu}\, \,q_{\mu}^{\dagger(3)}q_{\nu}^{\dagger(0)}
+V_{\mu\nu} \,\,q_{\mu}^{\dagger(1)}q_{\nu}^{\dagger(2)}\]
Which results in the effective meson decay Hamiltonian after some
manipulation
\beann
H_{m}=-\Phi_{\alpha}^{*\mu\lambda}\Phi_{\gamma}^{*\rho\sigma}\Phi_{\beta}^{\rho\lambda}
\,\,V_{\mu\nu} \,\,
m_{\alpha}^{\dagger}m_{\gamma}^{\dagger}m_{\beta}.
\eeann
We now consider the transition $m_\beta \rightarrow m_\alpha ^{\dagger
}m_\gamma $, where $\left| i\right> =m_\beta ^{\dagger }\left| 0\right> $
and $\left| f\right> =m_\alpha ^{\dagger }m_\gamma ^{\dagger }\left|
0\right>$.  Finally we find
\[
\left< f\right| H_{m}\left| i\right> =-\Phi _\alpha ^{*\rho \nu }\Phi
_\beta ^{\rho \lambda }\Phi _\gamma ^{*\mu \lambda }V_{\mu \nu }-\Phi
_\alpha ^{*\mu \lambda }\Phi _\beta ^{\rho \lambda }\Phi _\gamma ^{*\rho \nu
}V_{\mu \nu }
\]
The meson wave function is defined as 
\[
\Phi _\alpha ^{\mu \nu }=\chi _{S_\alpha ,M_\alpha ^S}^{S_1S_2} { }%
f_{f_\alpha ,M_\alpha ^f}^{ { }f_1f_2} { }C_{C_\alpha }^{C_1C_2}%
 { }\Phi _{\stackrel{\rightarrow }{P}_\alpha }^{\stackrel{\rightarrow }{p%
}_i\stackrel{\rightarrow }{p}_j}
\]
where $\chi$ is spin; $f$ is flavor and $C$ are color coefficients.
The spatial parte is given by
\[
\Phi _{\stackrel{\rightarrow }{P}_\alpha }^{\stackrel{\rightarrow }{p}_i%
\stackrel{\rightarrow }{p}_j}=\delta ^3(\stackrel{\rightarrow }{ { }%
P_\alpha }- { }\stackrel{\rightarrow }{p_i}-\stackrel{\rightarrow }{p_j}%
)\left( \frac 1{\pi b^2}\right) ^{3/4}e^{-\frac 1{8b^2}(\stackrel{%
\rightarrow }{p_i}-\stackrel{\rightarrow }{p_j})^2}
\]
We analyze the particular case $\rho^{+}( +\hat{z}) \rightarrow 
\pi ^{+}\pi ^{-}$ decay. For this decay have that
the flavor part is given by
$
\left| \rho ^{+}\right\rangle = | \pi ^{+}\rangle =-| u\bar{d}\rangle$ ;
$
| \pi ^0\rangle = \frac 1{\sqrt{2}}[ | u\bar{u}
\rangle - | d\bar{d}\rangle ]  
$
and the spin part is given by
$
| \rho ( +\hat{z}) \rangle =|\uparrow \stackrel{\_}{\uparrow }\rangle
$
;
$
| \pi \rangle =\frac 1{\sqrt{2}}[ | \uparrow \stackrel{%
\_}{\downarrow }\rangle -| \downarrow \stackrel{\_}{\uparrow }%
\rangle ] .
$
%
%
%
The transition matrix element is given by
\[
\left\langle  { }f { }\right| H_{FT}\left| i { }\right\rangle
=\delta ^3(\stackrel{\rightarrow }{ { }P_\beta }- { }\stackrel{%
\rightarrow }{P_\alpha }-\stackrel{\rightarrow }{P_\gamma }) { }h_{fi}
\]
with
\[
h_{fi}=\frac{2^3}{3^3}\gamma \left( \frac 1{\pi ^{3/4}b^{1/2}}\right) \frac Pb e^{-\frac{P^2}{12b^2}}
\]
This $h_{fi}$ decay amplitude can be combined with relativistic phase
space to give the differential decay rate, which is \cite{barnes}
\[
\frac{d\Gamma _{A\rightarrow BC}}{d\Omega }=2\pi \frac{PE_BE_C}{M_A}\left|
h_{fi}\right| ^2
\]
resulting
\[
\Gamma _{A\rightarrow BC}=\sqrt{\pi}\,\frac{2^7}{3^6}\left( \frac
g{2m_q}\right) ^2 M_\rho \left( \frac Pb\right) ^3 e^{-%
\frac{P^2}{6b^2}}.
\]
\begin{figure}
  \includegraphics[height=.3\textheight]{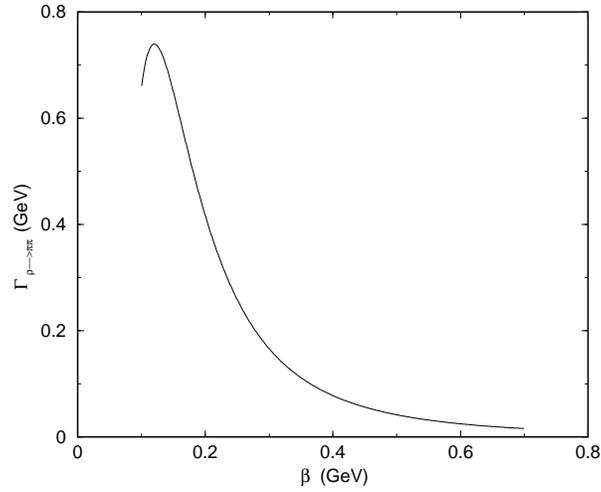}
  \caption{$\Gamma_{\rm theor}$ for $\rho\raw\pi\pi$;   
 $\Gamma_{\rm exp}=151$ MeV is obtained for $\beta =0.397 GeV$      }
\end{figure}
\section{Conclusions}
The Fock-Tani formalism is proven appropriate not only for hadron scattering but for decay.
The example decay process $\rho^{+}\rightarrow \pi^{+}\pi^{0}$ in the Fock-Tani formalism 
reproduces the predictions in the $^{3}P_{0}$ model.
The same procedure can be used for $f_{0}\left(M\right)$ and for heavier scalar mesons 
and compared with similar calculations with glue content.
\begin{theacknowledgments}
The authors acknowledges support from the Funda\c c\~ao de Amparo \`a Pesquisa 
do Estado do Rio Grande do Sul - FAPERGS. 
D.T.S. acknowledges support from the Coordena\c c\~ao de 
Aperfei\c coamento de Pessoal de N\'{\i}vel Superior - CAPES.
M.L.L.S. acknowledges support 
from the Conselho Nacional de Desenvolvimento Cient\'{\i}fico e Tecnol\'ogico
- CNPq. 
\end{theacknowledgments}

\end{document}